\documentstyle[11pt,epsf]{article}

\title{\large \bf 
COOPERATIVE JAHN-TELLER EFFECT \\ 
ON THE MAGNETIC STRUCTURE OF MANGANESE OXIDES
\footnote{To appear in proceedings of the conference 
``Science and Technology of Magnetic Oxides '99'', 
La Jolla, July 5--7, 1999.}
}

\author{Takashi Hotta, Seiji Yunoki, and Elbio Dagotto  \\ \\ 
National High Magnetic Field Laboratory  \\ 
Florida State University, Tallahassee, FL 32306}

\date{}

\begin{document}
\maketitle

\begin{abstract}
The magnetic structure of LaMnO$_3$ is investigated 
on three-dimensional clusters of MnO$_6$ octahedra
by using a combination of relaxation and Monte Carlo techniques.
It is found that the cooperative Jahn-Teller phonons 
lead to the stabilization of A-type antiferromagnetic 
and C-type orbital structures
in the physically relevant region of parameter space for LaMnO$_3$
with small corrections due to tilting effects.
The results suggest that 
strong Coulomb interactions are not necessary for a
qualitative description of undoped manganites.
In fact, it is shown that the present result is not essentially changed 
even if the Coulomb interaction is explicitly included. 
\end{abstract}

\section{Introduction}

The study of manganese oxides is receiving considerable attention 
in recent years both in its theoretical and experimental aspects
\cite{review}.
From the technological viewpoint, these materials could be used 
in the preparation of high-sensitive magnetic-field sensors due to 
their colossal magnetoresistance (CMR) phenomena. 
In addition, researchers in the condensed matter field have been  
interested in the rich phase diagram of these materials
originating from the competition and interplay 
among charge, spin, and orbital degrees of freedoms.
Obtaining a unified picture for this rich
phase diagram is a challenging open problem.

A prototype for the theoretical investigation of manganese oxides is 
the double-exchange (DE) framework, describing the hopping motion of 
e$_{\rm g}$-electrons ferromagnetically coupled to localized 
t$_{\rm 2g}$-spins. 
This idea has conceptually explained the appearance of 
ferromagnetism when holes are doped \cite{DEmodel}.
In addition, within the one orbital model, 
the existence of phase separation has been recently unveiled with the use of 
modern numerical techniques \cite{Yunoki}, leading to a potential
explanation of the CMR effect \cite{Moreo}.

However, in order to understand the fine details of the phase diagram of 
manganites, the one-orbital model is not sufficient since the highly 
nontrivial A-type spin antiferro (AF) and C-type orbital structures 
observed experimentally
in the undoped material LaMnO$_3$\cite{LaMnO3} cannot be properly addressed 
in such a simple context.
Certainly two-orbital models are needed to consider the nontrivial state
of undoped manganites. 
In this framework the two-band model without phonons has been studied before, 
and the importance of the strong Coulomb repulsion has been remarked for the
appearance of the A-AF state \cite{Coulomb,Mizokawa,Ishihara,Maezono}.
This is based upon the belief that the competition between kinetic 
and strong correlation effect determines the optimal orbital for 
e$_{\rm g}$-electrons and the lattice will be simply distorted to 
reproduce such optimal orbitals.
However, Coulombic approaches have presented conflicting results regarding the
orbital order that coexists with the A-type spin state, with several
approaches predicting G-type orbital order, which is not observed in practice.

While it is certainly correct that the orbital degrees of freedom 
play an essential role for the stabilization of A-AF, it should be 
noticed that our understanding is still incomplete.
In particular, it is important to remark that the orbital 
structure is tightly related to the Jahn-Teller (JT) distortion of 
MnO$_6$ octahedron. 
If each JT distortion would occur independently, optimal orbitals can be 
determined by minimizing the kinetic and interaction energy
of e$_{\rm g}$-electrons.
However, oxygens are shared between adjacent MnO$_6$ octahedra, 
indicating that the JT distortions occurs cooperatively. 
Especially in the undoped situation, all MnO$_6$ octahedra exhibit JT 
distortion, indicating that the cooperative effect is very important, 
as discussed by Kanamori \cite{Kanamori}.
Thus, in order to understand the magnetic and orbital structures in LaMnO$_3$, 
it is indispensable to optimize simultaneously the electron and lattice 
systems.
However, not much effort has been devoted to the microscopic treatment of 
the cooperative effect \cite{Allen},
although the JT effect in the manganese oxide
has been studied by several groups
\cite{Bandcal,JTpolaron,Koizumi,Yunoki2,Benedetti}.
Then, in the present work, a careful investigation of 
this problem is performed with some numerical techniques, 
focusing on $n=1$, where $n$ is the electron number per site.

In this paper, the optimal oxygen positions are 
determined by a relaxation technique 
to obtain the lattice distortions corresponding to several 
t$_{\rm 2g}$-spin magnetic structures.
In addition, Monte Carlo (MC) simulations were also performed to 
investigate the spin and orbital structure without {\it a priori}
assumptions for their order. 
It is found that A-AF, as well as the C-type orbital structure, 
occurs in realistic parameter regions for LaMnO$_3$, i.e., 
large Hund coupling between the e$_{\rm g}$-electron and t$_{\rm 2g}$-spin,
small AF interaction between t$_{\rm 2g}$-spins,
and strong electron-lattice coupling.
It should be emphasized that our results are obtained 
without the Coulomb interaction.
It is shown in a simple case that 
the optimized results are essentially unchanged even if the Coulomb 
interaction is included explicitly in the model.

The organization of this paper is as follows.
Section 2 is devoted to the formulation to include the cooperative effect
in the two-orbital model tightly coupled to the JT distortion, 
and some technical points are briefly discussed.
In Sec.~3, the results on the magnetic and orbital structures are 
provided and it is shown that 
the region of A-AF in the magnetic phase diagram
is reasonable for LaMnO$_3$, since the couplings needed agree with
experiments.
In Sec.~4, a prescription to obtain the C-type orbital order 
with the alternation of $3x^2-r^2$ and $3y^2-r^2$ orbitals 
is provided. 
Finally, in Sec.~5, the effect of the Coulomb interaction on the 
lattice distortion is discussed in the ferromagnetic state.
Throughout this paper, units such that $\hbar=k_{\rm B}=1$ are used.

\section{Formulation}

\subsection{Hamiltonian}

Let us consider the motion of e$_{\rm g}$-electrons tightly coupled to the 
localized t$_{\rm 2g}$-spins and the local distortions of the 
MnO$_6$ octahedra.
This situation is well described by
\begin{equation}
  \label{Hamiltonian}
  H=H_{\rm 2orb} + H_{\rm AFM} + H_{\rm el-ph} + H_{\rm el-el}.
\end{equation}
Here the first term indicates the two-orbital Hamiltonian, 
given by,
\begin{eqnarray}
 H_{\rm 2orb} = -\sum_{{\bf ia}\gamma \gamma' \sigma}
  t^{\bf a}_{\gamma \gamma'} c_{{\bf i} \gamma \sigma}^{\dag} 
  c_{{\bf i+a} \gamma' \sigma}
  - J_{\rm H} \sum_{{\bf i}\gamma\sigma \sigma'}
  {\bf S}_{\bf i} \cdot c^{\dag}_{{\bf i} \gamma \sigma}
  {\bf{\sigma}}_{\sigma \sigma'} c_{{\bf i} \gamma \sigma'},
\end{eqnarray}
where $c_{{\bf i}a \sigma}$ ($c_{{\bf i} b \sigma}$) is
the annihilation operator for an e$_{\rm g}$-electron with spin $\sigma$ 
in the $d_{x^2-y^2}$ ($d_{3z^2-r^2}$) orbital at site ${\bf i}$.
The vector connecting nearest-neighbor sites is ${\bf a}$, 
$t^{\bf a}_{\gamma \gamma'}$ is the hopping amplitude between $\gamma$- and 
$\gamma'$-orbitals connecting nearest-neighbor sites 
along the ${\bf a}$-direction via the oxygen 2$p$-orbital,
$J_{\rm H}$ is the Hund coupling,
${\bf S}_{\bf i}$ the localized classical t$_{\rm 2g}$-spin 
normalized to $|{\bf S}_{\bf i}|=1$,
and ${\bf {\sigma}}=(\sigma_1, \sigma_2, \sigma_3)$ are the Pauli matrices.

The second term is needed to account for the AFM character of the manganese
oxide, given by 
\begin{eqnarray}
  H_{\rm AFM} = J' \sum_{\langle {\bf i,j} \rangle}  
  {\bf S}_{\bf i} \cdot {\bf S}_{\bf j},
\end{eqnarray}
where $J'$ is the AF-coupling between nearest-neighbor t$_{\rm 2g}$-spins.

In the third term, the coupling of e$_{\rm g}$-electrons to
the distortion of MnO$_6$ octahedron is considered as 
\begin{eqnarray}
  H_{\rm el-ph} &=& g \sum_{{\bf i} \sigma\gamma\gamma'}
  c_{{\bf i} \gamma \sigma}^{\dag}
  (Q_{1{\bf i}} \sigma_0 +Q_{2{\bf i}} \sigma_1 + Q_{3{\bf i}}\sigma_3)
  _{\gamma\gamma'} c_{{\bf i} \gamma' \sigma} \nonumber \\ 
  &+& (1/2) \sum_{\bf i}[k_{\rm br} Q_{1{\bf i}}^2
  +k_{\rm JT}(Q_{2{\bf i}}^2+Q_{3{\bf i}}^2)], 
\end{eqnarray}
where $g$ is the electron-phonon coupling constant, 
$Q_{1{\bf i}}$ denotes the distortion for the breathing mode
of the MnO$_6$ octahedron, 
$Q_{2{\bf i}}$ and $Q_{3{\bf i}}$ are, respectively, 
JT distortions for the $(x^2-y^2)$- and $(3z^2-r^2)$-type modes,
and $\sigma_0$ is the $2 \times 2$ unit matrix.
Spring constants for breathing- and JT-modes are
denoted by $k_{\rm br}$ and $k_{\rm JT}$, respectively.

The final term indicates the Coulomb interactions between 
e$_{\rm g}$-electrons. As mentioned in Sec.~1, 
since only the JT phonons are investigated, 
the Coulomb interactions are neglected.
This point will be discussed later in the text (Sec.~5) 
where the explicit form of $H_{\rm el-el}$ is provided and briefly analyzed.

\subsection{Lattice distortion}

\begin{figure}[t]
\vskip1.5truein
\hskip-0.8truein
\centerline{\epsfxsize=3truein \epsfbox{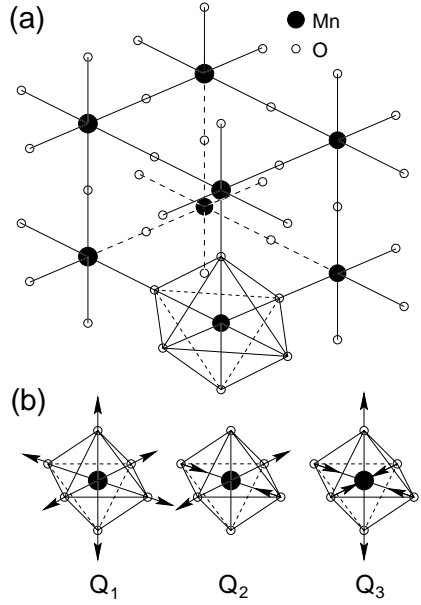} }
\vskip-1.0truein
\caption{(a) $2\times 2\times 2$ lattice composed of eight MnO$_6$
octahedra. Note that oxygens are shared by adjacent octahedra.
(b) Three kinds of distortions of octahedron considered in this paper.
The arrows indicate the direction of displacement of oxygens.}
\end{figure}

As shown in Fig.~1(a), oxygens are shared between adjacent octahedra, 
indicating that the local lattice distortions cannot be treated 
independently and a cooperative analysis is needed for this problem.
For this purpose, 
the normal coordinates for distortions of the MnO$_6$ 
octahedron, shown in Fig.~1(b), are written as \cite{Allen}
\begin{equation}
Q_{1 {\bf i}}=(1/\sqrt{3})(\Delta_{\bf xi}+ \Delta_{\bf yi}+\Delta_{\bf zi}),
\end{equation}
for the breathing mode, 
\begin{equation}
\label{q2}
Q_{2 {\bf i}}=(1/\sqrt{2})(\Delta_{\bf xi}- \Delta_{\bf yi}),
\end{equation}
and 
\begin{equation}
\label{q3}
Q_{3 {\bf i}}=(1/\sqrt{6})(2 \Delta_{\bf zi}- \Delta_{\bf xi}-\Delta_{\bf yi}),
\end{equation}
for the JT modes, where $\Delta_{\bf ai}$ is given by
\begin{equation}
  \Delta_{\bf ai} = \Delta_{\bf a}+\delta_{\bf ai}.
\end{equation}
The first term indicates the deviation from the cubic lattice,
given by $\Delta_{\bf a}=L_{\bf a}-L$, 
where $L_{\bf a}$ is the length between Mn-ions along the ${\bf a}$-axis
and $L=(L_{\bf x}+L_{\bf y}+L_{\bf z})/3$.
The second term is the contribution from the shift of oxygen position,
expressed by $\delta_{\bf ai}=u_{\bf i}^{\bf a}-u_{\bf i-a}^{\bf a}$,
where $u_{\bf i}^{\bf a}$ is the deviation of oxygen from the equilibrium 
position along the Mn-Mn bond in the ${\bf a}$-direction. 
By this consideration, the cooperative JT distortion as well as the 
macroscopic lattice deformation is reasonably taken into account.
Note that the buckling and rotational modes 
of MnO$_6$ octahedron are not explicitly included in this work. 
In general, $L_{\bf a}$ can be different for each direction,
depending on the bulk properties of the lattice.
Since the present work focuses on the microscopic mechanism for A-AF 
formation in LaMnO$_3$, 
the undistorted lattice with $L_{\bf x}=L_{\bf y}=L_{\bf z}$ 
is treated first, and then corrections will be added.

\subsection{Hopping amplitudes and energy scale}

In the cubic undistorted lattice, the hopping amplitudes are given by
\cite{Slater}
\begin{equation}
t_{\rm aa}^{\bf x}=-\sqrt{3}t_{\rm ab}^{\bf x}
=-\sqrt{3}t_{\rm ba}^{\bf x}=3t_{\rm bb}^{\bf x}=t,
\end{equation}
for the ${\bf x}$-direction, 
\begin{equation}
t_{\rm aa}^{\bf y}=\sqrt{3}t_{\rm ab}^{\bf y}=\sqrt{3}t_{\rm ba}^{\bf y}=
3t_{\rm bb}^{\bf y}=t,
\end{equation}
for the ${\bf y}$-direction, and 
\begin{equation}
t_{\rm bb}^{\bf z}=4t/3, 
t_{\rm aa}^{\bf z}=t_{\rm ab}^{\bf z}=t_{\rm ba}^{\bf z}=0,
\end{equation}
for the ${\bf z}$-direction.
Throughout this paper, the energy unit is $t$.

Corresponding to this choice of the energy unit, 
the length in the lattice distortion is scaled by $\sqrt{t/k_{\rm JT}}$.
As a result of this scaling, 
a non-dimensional electron-phonon coupling constant
$\lambda$ is defined as 
\begin{equation}
 \lambda = g/\sqrt{k_{\rm JT}t}.
\end{equation}
It is noted that this coupling constant can be related to the static 
JT energy, 
which is conventionally defined by $E_{\rm JT}=g^2/(2k_{\rm JT})$, as
\begin{equation}
 E_{\rm JT} = t\lambda^2/2.
\end{equation}
Note also that the present length scale is rewritten as
\begin{equation}
 \sqrt{t/k_{\rm JT}}= \ell_{\rm JT}/\lambda,
\end{equation}
where $\ell_{\rm JT}=g/k_{\rm JT}$ is the characteristic length
for the JT distortion.
From the experimental result, $\ell_{\rm JT}$ is estimated as 
$0.3$\AA~\cite{LaMnO3}, which is a typical length in this context.

As for the spring constant for the breathing mode, 
it is expressed as $k_{\rm br}=\beta k_{\rm JT}$ and the ratio $\beta$
is treated as a parameter. 
If it is plausibly assumed that the reduced masses for those modes are equal,
this ratio is given by $\beta=(\omega_{\rm br}/\omega_{\rm JT})^2$,
where $\omega_{\rm br}$ and $\omega_{\rm JT}$ are the vibration energies 
for manganite breathing- and JT-modes, respectively. 
From experimental results and band-calculation data,
$\omega_{\rm br}$ and $\omega_{\rm JT}$ are, respectively, 
estimated as $\sim 700$cm$^{-1}$ and $500$-$600$cm$^{-1}$\cite{Iliev}.
Then, throughout this work, $\beta$ is taken as $2$,
although the results presented here are basically unchanged as long as 
$\beta$ is larger than unity.

In this work, the change in $t$ due to the displacement of oxygen position
is not taken into account, but such an effect is shown to 
be very small as follows.
Due to the pseudo-potential theory \cite{Harrison},
the exact hopping amplitude, for example, along
the ${\bf x}$-direction between a-orbitals in ${\bf i}$
and ${\bf i+x}$ sites, is expressed as 
\begin{equation}
 t_{\rm aa}^{\bf x} = {t \over (1-\epsilon^2)^{7/2}},
\end{equation}
with $\epsilon=|2u_{\bf i}^{\bf x}|/L_{\bf x}$.
It should be remarked that the change in $t_{\rm aa}^{\bf x}$ due to 
the oxygen shift is 
of the order of $\epsilon^2$, not of the order of $\epsilon$.
Since $\epsilon$ is estimated to be at most a few percent,
the change is considered to be negligible.
Thus, in the present work, such a change is not included to 
avoid unnecessary complication in the calculation. 
However, when the distorted lattice is considered, namely, 
when the deviation from $L_{\bf x}=L_{\bf y}=L_{\bf z}$ is taken into 
account, the change in the hopping matrix due to this distortion 
should be included, because the effect is of the order 
of $\Delta_{\bf a}/L$
in this case, not of the order of $(\Delta_{\bf a}/L)^2$
This point will be discussed again in Sec.~4.

\subsection{Techniques}

\begin{figure}[t]
\ \vskip-1truein
\hskip-3.2truein
\centerline{\epsfxsize=2.8truein \epsfbox{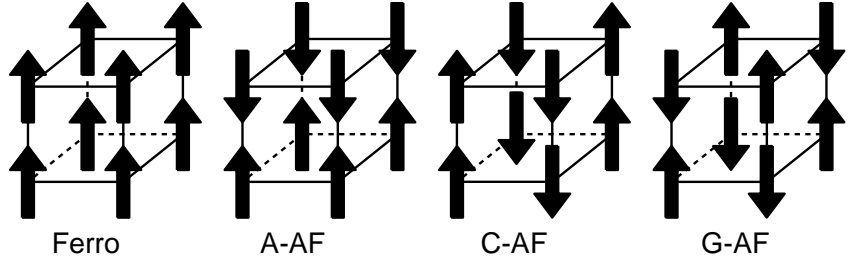} }
\vskip-6.0truein
\caption{Magnetic structures in $2\times 2 \times 2$ clusters.}
\end{figure}

To study Hamiltonian Eq.~(\ref{Hamiltonian}) without $H_{\rm el-el}$, 
two numerical techniques have been applied.
One is the relaxation technique, in which the optimal positions of the 
oxygens are determined by minimizing the total energy.
In this calculation, only the stretching mode for the octahedron, 
is taken into account.
Moreover, the relaxation has been performed for fixed structures of 
the t$_{\rm 2g}$-spins such as ferro (F), A-type AF (A-AF), 
C-type AF (C-AF), and G-type AF (G-AF), shown in Fig.~2.
The advantage of this method is that the optimal orbital structure
can be rapidly obtained on small clusters.
However, the assumptions involved in the relaxation procedure
should be checked with an independent method.

Such a check is performed with the unbiased MC simulations
used before for one- and two-dimensional 
clusters using non-cooperative JT-phonons \cite{Yunoki2}.
The dominant magnetic and orbital structures
are deduced from correlation functions.
In the MC method, the clusters currently reachable are
$2 \times 2 \times 2$, $4 \times 4 \times 2$, and 
$4\times 4 \times 4$.  In spite of this size limitation,
arising from the large number of degrees of 
freedom in the problem, the available clusters are sufficient for our
mostly qualitative purposes.
In addition, the remarkable agreement between MC and relaxation methods 
lead us to believe that our results 
are representative of the bulk limit.

\section{Results}

\subsection{Magnetic structure}

\begin{figure}[t]
\ \vskip0.6truein
\hskip-0.5truein
\centerline{\epsfxsize=5truein \epsfbox{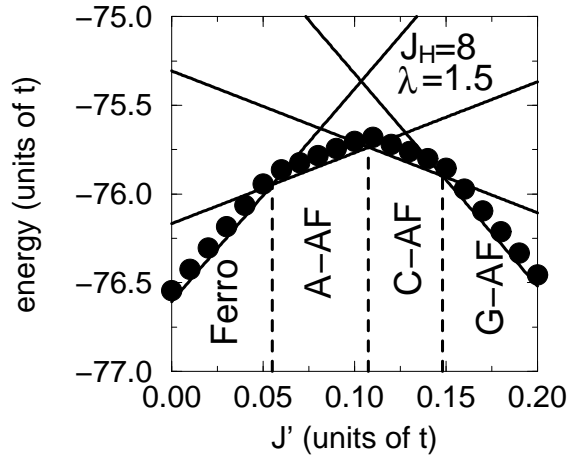} }
\vskip-2.5truein
\caption{
Total energy as a function of $J'$ on a
$2\times 2 \times 2$ lattice with $J_{\rm H}=8$ and $\lambda=1.5$.
The solid lines and circles indicate the relaxation and MC results,
respectively. MC simulations have been performed at temperature $1/200$.}
\end{figure}

In Fig.~3, the mean-energy is presented as a function of $J'$
for $J_{\rm H}=8$ and $\lambda=1.5$, on a $2 \times 2 \times 2$ cluster 
with open boundary conditions.
The solid lines and symbols indicate the results obtained with the 
relaxation technique and MC simulations, respectively.
The agreement is excellent, showing that the
relaxation method is accurate. 
The small deviations between the results of the two techniques are caused
by temperature effects.
As intuitively expected, with increasing $J'$ the optimal magnetic 
structure changes from ferro- to antiferromagnetic, and this
occurs in the order F$\rightarrow$A-AF$\rightarrow$C-AF$\rightarrow$G-AF.

To check size effects, the t$_{\rm 2g}$-spin
correlation function $S({\bf q})$ was calculated also in 
$4 \times 4 \times 2$ and $4\times 4 \times 4$ clusters,
where 
\begin{equation}
S({\bf q})=
(1/N)\sum_{\bf i,j}e^{-i{\bf q}\cdot({\bf i}-{\bf j})}
\langle {\bf S}_{\bf i}\cdot{\bf S}_{\bf j} \rangle.
\end{equation}
Here $N$ is the number of sites and $\langle \cdots \rangle$
indicates the thermal average value.
As shown in Fig.~4, with increasing $J'$ the dominant correlation 
changes in the order of ${\bf q}=(0,0,0)$, $(\pi,0,0)$, $(\pi,\pi,0)$, 
and $(\pi,\pi,\pi)$.
The values of $J'$ at which the spin structures changes
agree well with those in Fig.~3.

\begin{figure}[t]
\ \vskip0.6truein
\hskip-0.5truein
\centerline{\epsfxsize=5truein \epsfbox{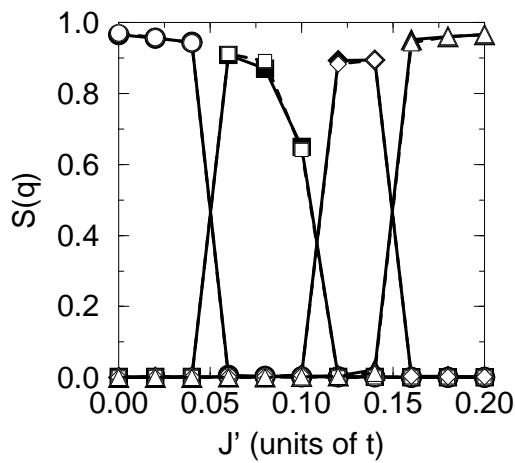} }
\vskip-2.5truein
\caption{
Spin correlation function $S({\bf q})$ obtained by
MC simulations as a function of $J'$, at $J_{\rm H}=8$ and $\lambda=1.5$.
Solid and open symbols denote the results
in $4\times 4 \times 2$ and $4\times 4 \times 4$ clusters, respectively.
Circles, squares, diamonds, and triangles indicates $S({\bf q})$ for
${\bf q}=(0,0,0)$, $(\pi,0,0)$, $(\pi,\pi,0)$,
and $(\pi,\pi,\pi)$, respectively.}
\end{figure}

\subsection{Orbital structure}

The shapes of the occupied orbital arrangement with the lowest energy 
for F, A-AF, C-AF, and G-AF magnetic structures are shown in Fig.~5. 
For the F-case, the G-type orbital structure is naively expected,
because it is believed that the ferromagnetic spin structure
is favored by the AF orbital configuration. 
However, a more complicated orbital structure is stabilized in the actual
calculation, 
indicating the importance of the cooperative treatment for JT-phonons.
For the A-AF state,
only the C-type structure is depicted in Fig.~5, but the 
G-type structure, obtained by a $\pi$/2-rotation of the upper $x$-$y$ plane
of the C-type state, was found to have {\it exactly} the same energy.
Small corrections will remove this degeneracy in favor of the C-type as
described in the next section.
For C- and G-AF, the obtained orbital structures are 
G- and C-types, respectively.

Note that there exists an additional triplet degeneracy 
in the A-type state due to the cubic symmetry for each magnetic structure: 
If axes are changed cyclically
($x \rightarrow y, y \rightarrow z, z \rightarrow x$),
the optimized orbital structure is also transformed by 
this cyclic change, but the energy is invariant.
Then, the magnetic and orbital structure in LaMnO$_3$ 
occurs through a {\it spontaneous} symmetry breaking process.

Although the same change of the magnetic structure due to $J'$ was 
already reported in the electronic model with purely Coulomb 
interactions \cite{Maezono},
the orbital structures in those previous
calculations were G-, G-, A-, and A-type for the
F-, A-AF, C-AF, and G-AF spin states, respectively.
Note that for the A-AF state, of relevance for the undoped manganites,
the G-type order was obtained \cite{Maezono},
although in another treatment for the Coulomb interaction,
the C- and G-type structures were found to be degenerate \cite{Mizokawa},
as in our calculation. 
Thus, the stabilization in experiments of the C-type orbital structure is
still puzzling both in the JT and Coulomb mechanisms.
This point will be discussed later in the text.

\begin{figure}[t]
\vskip-1.7truein
\hskip-0.35truein
\centerline{\epsfxsize=6.5truein \epsfbox{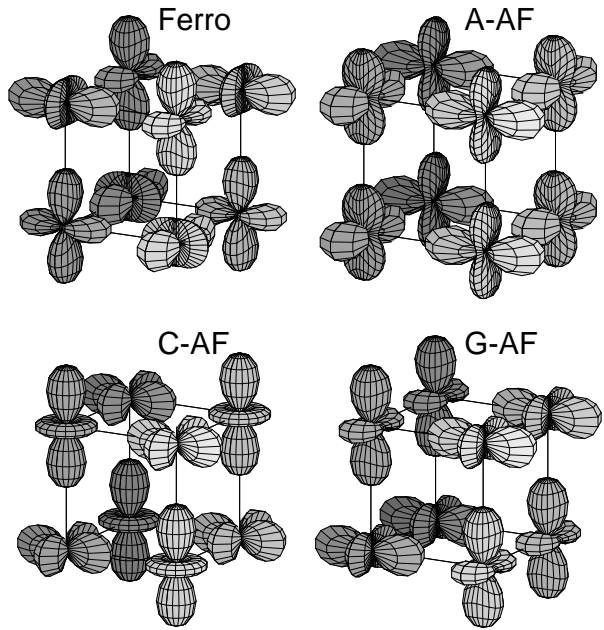} }
\vskip-2.8truein
\caption{Optimized orbital structure for each magnetic structure.}
\end{figure}

\subsection{Magnetic phase diagram}

In Figs.~6(a) and (b), the phase diagrams 
on the $(J', \lambda)$-plane are shown 
for $J_{\rm H}=4$ and $8$, respectively.
The curves are drawn by the relaxation method.
As expected, the F-region becomes wider with increasing $J_{\rm H}$.
When $\lambda$ is increased at fixed $J_{\rm H}$, 
the magnetic structure changes
from F$\rightarrow$A-AF$\rightarrow$C-AF$\rightarrow$G-AF.
This tendency is qualitatively understood if the two-site problem is 
considered in the limit $J_{\rm H} \gg 1$ and $E_{\rm JT} \gg 1$.
The energy-gain due to the second-order hopping process of 
e$_{\rm g}$-electrons 
is roughly $\delta E_{\rm AF} \sim 1/J_{\rm H}$ and 
$\delta E_{\rm F} \sim 1/E_{\rm JT}$ for AF- and F-spin pairs, respectively.
Increasing $E_{\rm JT}$, $\delta E_{\rm F}$ decreases,
indicating the relative stabilization of the AF-phase.
In our phase diagram, the A-AF phase appears for 
$\lambda \ge 1.1$ and $J'\le 0.15$.
This region does not depend much on $J_{\rm H}$, as long as 
$J_{\rm H} \gg 1$.
Although $\lambda$ seems to be large, it is 
realistic from an experimental viewpoint:
$E_{\rm JT}$ is $0.25$eV from photoemission experiments \cite{Shen}
and $t$ is estimated as $0.2 \sim 0.5$eV \cite{Saito},
leading to $1 \le \lambda \le 1.6$.
As for $J'$, it is estimated as $0.02 \le J' \le 0.1$ \cite{Ishihara,Perring}.
Thus, the location in parameter-space of the A-AF state found here
is reasonable when compared with experimental results for LaMnO$_3$.

\begin{figure}[t]
\hskip-0.2truein
\vskip0.5truein
\centerline{\epsfxsize=3.6truein \epsfbox{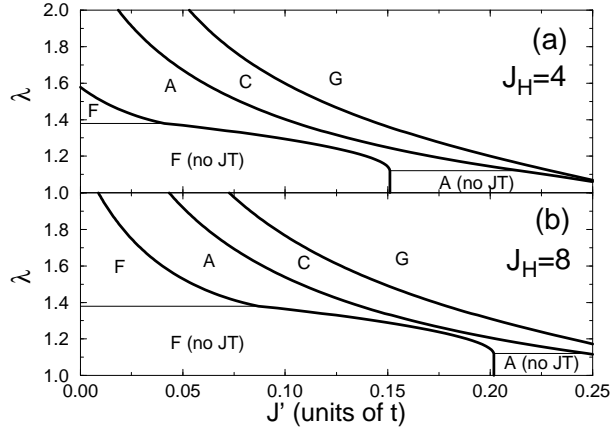} }
\vskip-1truein
\caption{Magnetic phase diagram on the $(J',\lambda)$ plane
for (a) $J_{\rm H}=4$ and (b) $8$ (relaxation method).
Below the thin solid lines in the F and A-AF regions,
the JT-distortion disappears, suggesting that the system becomes metallic.
}
\end{figure}

\section{Orbital order in the A-AF phase}

Let us now focus on the orbital structure in the A-AF phase.
In the cubic lattice studied thus far, the 
C- and G-type orbital structures are degenerate, and it is unclear
whether the orbital pattern in the $x$-$y$ plane corresponds
to the alternation of $3x^2-r^2$ and $3y^2-r^2$ orbitals 
observed in experiments \cite{LaMnO3}.
To remedy the situation, some empirical facts observed 
in manganites become important:
(i) The MnO$_6$ octahedra are slightly tilted from 
each other, leading to modifications in the hopping matrix.
Among these modifications, the generation of a
 non-zero value for $t_{\rm aa}^{\bf z}$ 
is important.
(ii) The lattice is not cubic, but the relation 
$L_{\rm in} > L_{\rm out}$ holds, where 
$L_{\rm in}=L_{\bf x}=L_{\bf y}$ and $L_{\rm out}=L_{\bf z}$.
From experimental results \cite{LaMnO3}, 
these numbers are estimated as $L_{\rm in}=4.12$\AA~and $L_{\rm out}=3.92$\AA,
indicating that the distortion with $Q_3$-symmetry occurs spontaneously.

\begin{figure}[t]
\ \vskip0.5truein
\hskip-0.2truein
\centerline{\epsfxsize=3.6truein \epsfbox{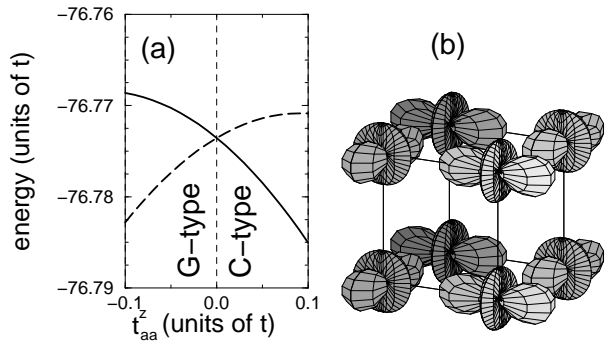} }
\vskip-1.5truein
\caption{(a) Total energy as a function of $t_{\rm aa}^{\bf z}$
on the $2\times 2 \times 2$ lattice with $L_{\bf x}=L_{\bf y}>L_{\bf z}$
for $J_{\rm H}=8$, $\lambda=1.6$
and $J'=0.05$ (relaxation method).
The solid and dashed curves denote the C- and
G-orbital states, respectively.
(b) Orbital structure in the A-AF phase
for $t_{\rm aa}^{\bf z}=0^{+}$ and $L_{\bf x}=L_{\bf y}>L_{\bf z}$.
These shapes are very close to purely $3x^2-r^2$ and $3y^2-r^2$ types.}
\end{figure}

\begin{figure}[t]
\ \vskip0.7truein
\hskip-1.2truein
\centerline{\epsfxsize=3.0truein \epsfbox{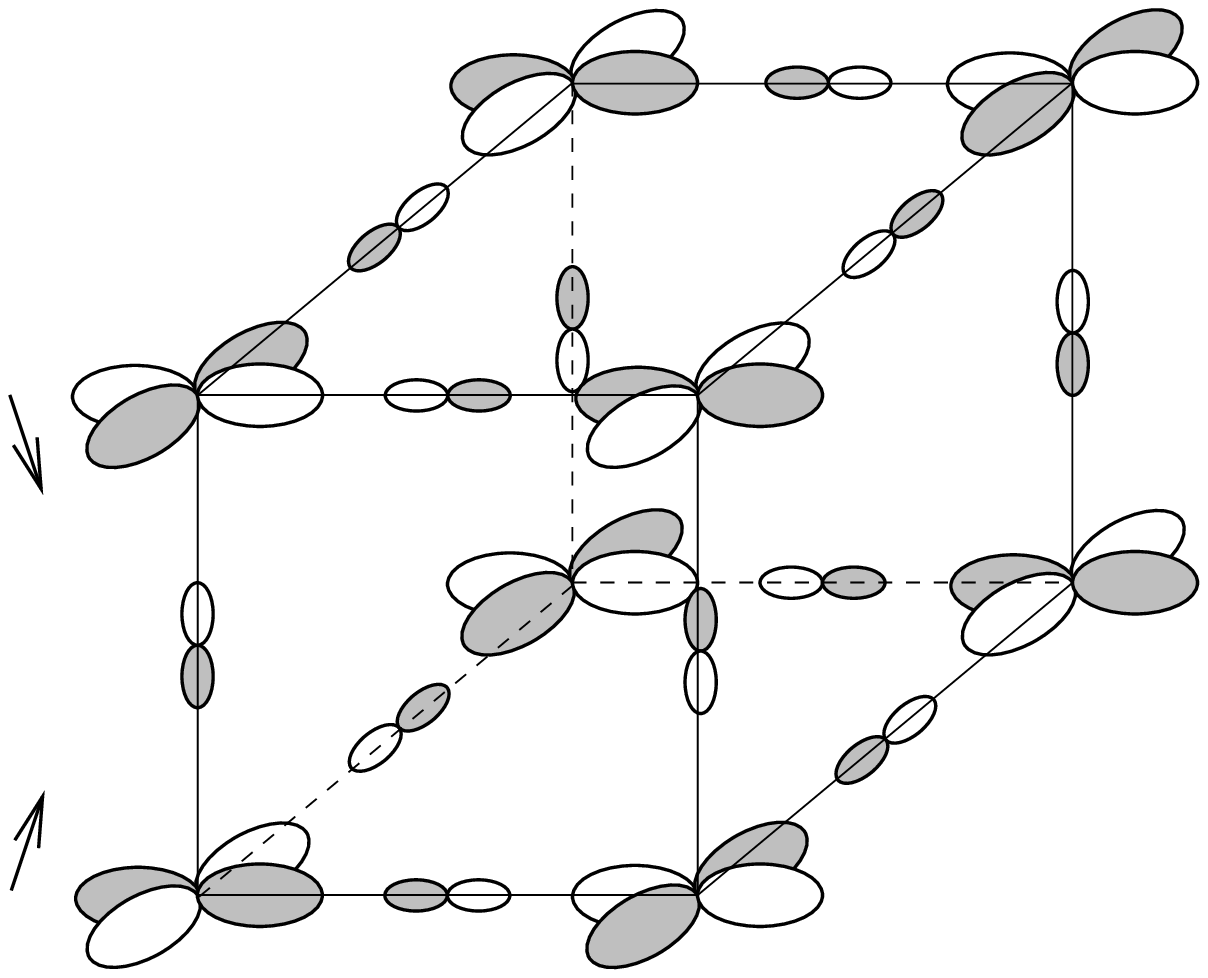} }
\vskip-1.7truein
\caption{Configuration of $x^2-y^2$ orbitals connected by oxygen $2p$
orbitals in a $2\times 2 \times 2$ cluster. 
The open and hatched parts indicate the plus and minus sign
in the orbitals, respectively.
Arrows indicate the direction of the tilting.}
\end{figure}

For the inclusion of the point (ii), $Q_{3{\bf i}}$ in Eq.~(\ref{q3}) 
is rewritten as 
\begin{equation}
Q_{3 {\bf i}}=Q_{3}^{(0)} 
+(1/\sqrt{6})(2 \delta_{\bf zi}- \delta_{\bf xi}-\delta_{\bf yi}),
\end{equation}
where $Q_{3}^{(0)}$ indicates the spontaneous distortion with $Q_3$-symmetry,
given by 
\begin{equation}
Q_{3}^{(0)} = \sqrt{2/3}(L_{\rm out}- L_{\rm in}).
\end{equation}
This length is rewritten in the non-dimensional form as
$Q_{3}^{(0)} = -\eta \lambda$, where $\eta$ is a numerical factor given by 
$\eta=\sqrt{2/3}(L_{\rm in}- L_{\rm out})/\ell_{\rm JT}$,
estimated as $0.5$ by using the experimental data.
Note that the hopping amplitude becomes different from those in the 
$x$-$y$ plane due to this distortion.
As discussed shortly in subsection 2.3,
it is obtained as $t_{\rm bb}^{\bf z}=(4t/3)(L_{\rm in}/L_{\rm out})^7$.
As for $J'$ along the ${\bf z}$-direction, 
it is given by $J'(L_{\rm in}/L_{\rm out})^{14}$,
since the superexchange interaction is proportional to the square of
the hopping amplitude.

Motivated by these observations, the energies for C- and G-type 
orbital structures were recalculated including this time a nonzero value
for $t_{\rm aa}^{\bf z}$ in the magnetic A-AF state (see Fig.~7a)).
In Fig.~8, the configuration of $x^2-y^2$ orbitals is depicted.
From this figure, it is intuitively understood that 
if the $x^2-y^2$ orbitals in the upper and lower planes are tilted 
from the $x$-$y$ plane as showing by arrows in the figure, 
there appears finite a hopping integral between adjacent $x^2-y^2$ orbitals 
along the $z$-direction and 
the sign of this hopping amplitude is the same as that of 
$t_{\rm aa}^{\bf x}$ or $t_{\rm aa}^{\bf y}$.
Thus, in the real material, the tilting of the MnO$_{6}$ 
octahedra will always lead to a {\it positive} value for $t_{\rm aa}^{\bf z}$
and the results of Fig.~7(a) 
suggest that the C-type orbital structure should 
be stabilized in the real materials.
The explicit shape of the occupied orbitals is shown in Fig.~7(b).
The experimentally relevant 
C-type structure with the approximate alternation of $3x^2-r^2$ and 
$3y^2-r^2$ orbitals is indeed successfully obtained by this procedure.
Although the octahedron tilting actually leads to a change of all hopping
amplitudes, effect not including in this work,
the present analysis is sufficient to show that
the C-type orbital structure is stabilized in the A-AF magnetic phase when
$t_{\rm aa}^{\bf z}$ is a small positive number, as it occurs
in the real materials.
Our investigations show that this mechanism to stabilize the C-type structure
works also for the purely electronic model in the Hartree-Fock approximation.

\section{Discussion and Summary}

In this work, the Coulomb interaction term $H_{\rm el-el}$ has been 
neglected, but this detail needs further clarification.
For this purpose, $H_{\rm el-el}$ is written as
\begin{eqnarray}
H_{\rm el-el} = U \sum_{{\bf i}\gamma} 
n_{{\bf i}\gamma\uparrow} n_{{\bf i}\gamma\downarrow}
+U' \sum_{{\bf i}\sigma \sigma'} 
n_{{\bf i}{\rm a}\sigma} n_{{\bf i}{\rm b}\sigma'}
+J \sum_{{\bf i}\sigma \sigma'}
c^{\dag}_{{\bf i}{\rm a}\sigma}c^{\dag}_{{\bf i}{\rm b}\sigma'}
c_{{\bf i}{\rm a}\sigma'}c_{{\bf i}{\rm b}\sigma},
\end{eqnarray}
where $U$ is the intra-orbital Coulomb interaction, 
$U'$ the inter-orbital Coulomb interaction,
and $J$ is the inter-orbital exchange interaction.
For Mn-oxides, they are estimated as $U=7$eV, $J=2$eV, and 
$U'=5$eV \cite{Ishihara}, which are large compared to $t$.
However, the result for the optimized distortion described in this
paper, obtained without the Coulomb interactions, is not expected 
to change,
since the energy gain due to the JT-distortion is maximized 
when a single e$_{\rm g}$-electron is present per site. This is 
essentially the same effect as produced by a short-range repulsion.
In fact, the MC simulations show that the probability of double
occupancy of a single orbital is negligible in the window of
couplings where the A-type spin and C-type orbital state is stable.

In order to confirm the above statement, 
the JT- and breathing-distortions were calculated as a function of $U'$ 
by using the Exact Diagonalization method on a $2 \times 2$ cluster 
in the F-state in which $U$ and $J$ can be neglected.
The result is shown in Fig.~9, where 
$Q_{\rm JT}$ and $Q_{\rm br}$ are defined as
\begin{equation}
Q_{\rm JT} = (1/N) \sum_{\bf i} \sqrt{ Q_{2{\bf i}}^2+Q_{3{\bf i}}^2},
\end{equation}
and 
\begin{equation}
Q_{\rm br} = (1/N) \sum_{\bf i} |Q_{1{\bf i}}|,
\end{equation}
respectively.
As expected, the mean value of the breathing-mode distortion is almost 
zero and only the JT-mode is active in the case of $\beta=2$.
It is noted that the dependence of $Q_{\rm JT}$ on $U'$ is very weak,
indicating that the optimized distortion is not affected by 
the Coulomb interaction.
The orbital arrangement in this $2 \times 2$ lattice is 
exactly the same as that in the $x$-$y$ plane of the orbital structure 
for the A-AF phase in Fig.~5
and this arrangement is unchanged by the inclusion of $U'$. 
Note also that $Q_{\rm JT}$ is gradually increased with the increase 
of $U'$, although the dependence is weak.
This suggests that the JT-distortion without $U'$ is reproduced at smaller 
value of $\lambda$ if $U'$ is included explicitly \cite{Benedetti}.
Thus, it is expected that 
the A-AF state will be stabilized at smaller $\lambda$ 
improving the comparison of our results with experiments.
Based on all these observations, it is believed that the effect of 
the Coulomb interaction is not crucial for the appearance of the A-AF
state with the proper orbital order. Another way to rationalize this
result is that the integration of the JT-phonons at large
$\lambda$ will likely induce Coulombic interactions dynamically.

\begin{figure}[t]
\ \vskip-0.5truein
\centerline{\epsfxsize=5.5truein \epsfbox{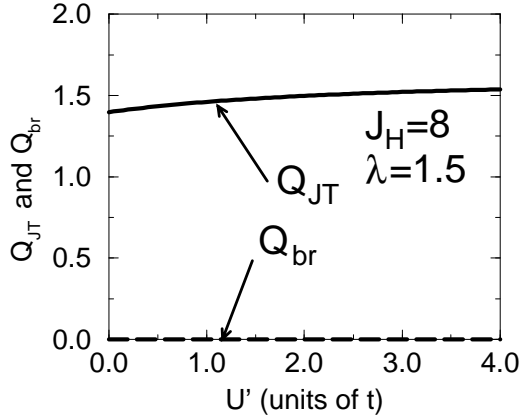} }
\ \vskip-2truein
\caption{Mean-value of the breathing- and JT-mode distortions
as a function of $U'$ for the F-phase on a $2\times 2$ cluster 
for $J_{\rm H}=8$ and $\lambda=1.5$.}
\end{figure}

Finally, let us briefly discuss transitions induced
by the application of external magnetic fields on undoped manganites.
When the A-AF state is stabilized, 
the energy difference (per site) obtained in our study
between the A-AF and F states is about $t/100$.
As a consequence, magnetic fields of $20 \sim 50$T 
could drive the transition from A-AF to F order
accompanied by a change of orbital structure, 
interesting effect which 
may be observed in present magnetic field facilities.

In summary, with the use of numerical techniques at $n=1$, 
it has been shown that the A-AF state is stable in models with
JT-phonons, using coupling values physically reasonable for LaMnO$_3$. 
Our results indicate that it is not necessary to include large Coulombic 
interactions in the calculations to reproduce experimental results for
the manganites.
Considering the small but important effect of the octahedra tilting of 
the real materials, 
the C-type orbital structure (with the alternation pattern of
$3x^2-r^2$ and $3y^2-r^2$ orbitals) has been successfully reproduced
for the A-AF phase in this context.

\section*{Acknowledgments}

T.H. is grateful to Y. Takada and H. Koizumi for enlightening discussion.
T.H. is supported from the Ministry of Education, Science,
Sports, and Culture of Japan. 
E.D. is supported by grant NSF-DMR-9814350.


\end{document}